\documentclass[twoside,10pt,a4paper]{newFNLstyle}
\usepackage{graphics}
\usepackage{cite}

\def\w0{\omega_0}

\begin{document}

\volnumpagesyear{3}{2}{000--000}{2003}
\dates{21 November 2002}{26 March 2003}{27 March 2003}

\title{NOISE INDUCED PHENOMENA IN LOTKA-VOLTERRA SYSTEMS}

\authorsone{B. Spagnolo$^*$, A. Fiasconaro and D. Valenti}
%\thanks{Use footnotes only to indicate if permanent and present addresses are different.
%Funding information should go in the Acknowledgement section.}
\affiliationone{Istituto Nazionale di Fisica della Materia,
Unit\`a di Palermo and Dipartimento di Fisica e Tecnologie
Relative, University of Palermo} \mailingone{Viale delle Scienze,
I-90128 Palermo, Italy \\ $^*$spagnolo@unipa.it}

%\authorstwo{FOURTH AUTHOR}
%\affiliationtwo{Full affiliation}
%\mailingtwo{and mailing address}

% Add author/affliation/mailing sets as required

\maketitle

\markboth{B. Spagnolo, A. Fiasconaro and D. Valenti}{Noise Induced
Phenomena in Lotka-Volterra Systems}

\pagestyle{myheadings}
% Comment this out to remove the running heads

% Keywords have to before the abstract I'm afraid.

\begin{abstract}
We study the time evolution of two ecosystems in the presence of
external noise and climatic periodical forcing by a generalized
Lotka-Volterra (LV) model. In the first ecosystem, composed by two
competing species, we find noise induced phenomena such as: (i)
quasi deterministic oscillations, (ii) stochastic resonance, (iii)
noise delayed extinction and (iv) spatial patterns. In the second
ecosystem, composed by three interacting species (one predator and
two preys), using a discrete model of the LV equations we find
that the time evolution of the spatial patterns is strongly
dependent on the initial conditions of the three species.
\end{abstract}

\section{Introduction}

The study of the effects of noise is a well established subject in
several different disciplines ranging from physics, to chemistry
and to biology \cite{Fre}. The essential role of the noise in
theoretical ecology however has been recently recognized. Some key
questions in population ecology are related to the understanding
of the role that noise, climatic forcing and nonlinear
interactions between individuals of the same or different species
play on the dynamics of the ecosystems
\cite{Zim,Com,Bjo,Sch,Sae,Ciu,Pas,Gia,Man,Dro,Cir}. More recently
noise induced effects on population dynamics have been
investigated \cite{Spa1,Vil,Bar,Spa2,Roz,Val}. In this paper we
study the dynamics of two and three interacting species in the
presence of multiplicative noise and a periodic driving force.
Specifically we consider: (a) two competing species and (b) three
interacting species. The multiplicative noise models the
interaction between the species and the environment, and the
driving force mimics the climatic temperature oscillations. In
case (a) the interaction parameter between the species is a
stochastic process which obeys an Ito stochastic differential
equation. The noise induces spatio-temporal patterns,
quasi-deterministic oscillations, stochastic resonance (SR)
phenomenon \cite{Gam,Man1} and noise delayed extinction, which is
akin of the noise enhanced stability \cite{Agu,Man2}. In case (b)
we find noise induced spatial patterns whose time evolution is
strongly dependent on the initial conditions. Specifically we
consider two different initial conditions: (i) an uniform initial
distribution which gives rise to anticorrelated spatio-temporal
patterns between the two preys, and (ii) a peaked initial
distribution which gives rise periodically to a strong correlation
between the patterns of the three species.

\section{The Model}
\subsection{Two competing species}
Time evolution of two competing species is obtained within the
formalism of the Lotka-Volterra equations~\cite{Lot} using the Ito
scheme

\begin{eqnarray}
\frac{dx}{dt}&=&\mu\thinspace x\thinspace(\alpha-x-\beta_x(t) y)+x\thinspace\xi_x(t),\\
\frac{dy}{dt}&=&\mu\thinspace
y\thinspace(\alpha-y-\beta_y(t)x)+y\thinspace\xi_y(t),
\label{LotVol}
\end{eqnarray}
where $\xi_i(t)$ is Gaussian white noise with zero mean and
$\langle \xi_i(t)$$\xi_j(t')\rangle$ = $\sigma
\delta(t-t')\delta_{ij}$. The multiplicative noise models the
interaction between the environment and the species. The
interaction parameters $\beta_x$ and $\beta_y$ are characterized
by a critical value corresponding to $\beta_c = 1$. We choose
$\beta_x = \beta_y = \beta$. For $\beta < \beta_c$ a coexistence
regime of the two species is established, while for $\beta >
\beta_c$ an exclusion regime takes place, i.e. in a finite time
one of the two species extinguishes.

\begin{figure}[htbp]
\centering{\resizebox{6cm}{!}{\includegraphics{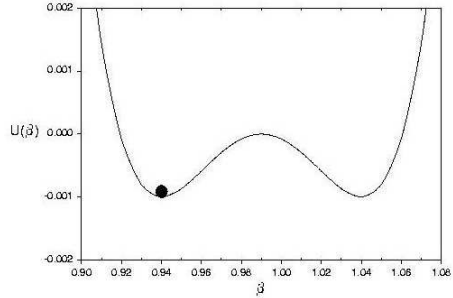}}}
%\centering{\frame{\rule{2cm}{2cm}}}
\caption{The bistable potential $U(\beta)$ of the interaction
parameter $\beta(t)$, centered at $\beta = 0.99$ (coexistence
region). The values of the parameters are $h=10^{-3}$,
$\rho=10^{-2}$, $\eta=0.05$. The initial value of $\beta(t) $ is
$\beta(0) = 0.94$ (bottom of left well).}
\end{figure}
The two regimes correspond to stable states of the deterministic
Lotka-Volterra model. It is then interesting to investigate the
time evolution of the ecosystem for $\beta$ varying around the
critical value $\beta_c$ in the presence of fluctuations, due to
the significant interaction with the environment. The interaction
parameter $\beta$ therefore can be described by a stochastic
process which obeys the following differential equation

\begin{equation}
\frac{d\beta(t)}{dt} = -\frac{dU(\beta)}{d\beta} + \gamma cos
\thinspace\w0 t + \xi_\beta(t),
\label{beta_eq}
\end{equation}
where $U(\beta)$ is a bistable potential (see Fig.1)

\begin{equation}
U(\beta) = h(\beta-(1+\rho))^4/\eta^4-2h(\beta-(1+\rho))^2/\eta^2,
\label{U(beta)}
\end{equation}
$\xi_{\beta}(t)$ is a Gaussian white noise with the usual
statistical properties: $\langle \xi_{\beta}(t)\rangle=0$ and
$\langle \xi_{\beta}(t)\xi_{\beta}(t')\rangle
=\sigma\delta(t-t')$. We perform simulations of Eqs.~(1) and
(\ref{LotVol}) by using a bistable potential $U(\beta)$ centered
at $\beta=0.99 < 1$, i.e. in the coexistence region. We consider
as initial value of the interaction parameter $\beta(0)=0.94$,
which corresponds to the left minimum of the potential well (see
Fig.1).
\begin{figure}[htbp]
\centering{\resizebox{13cm}{!}{\includegraphics{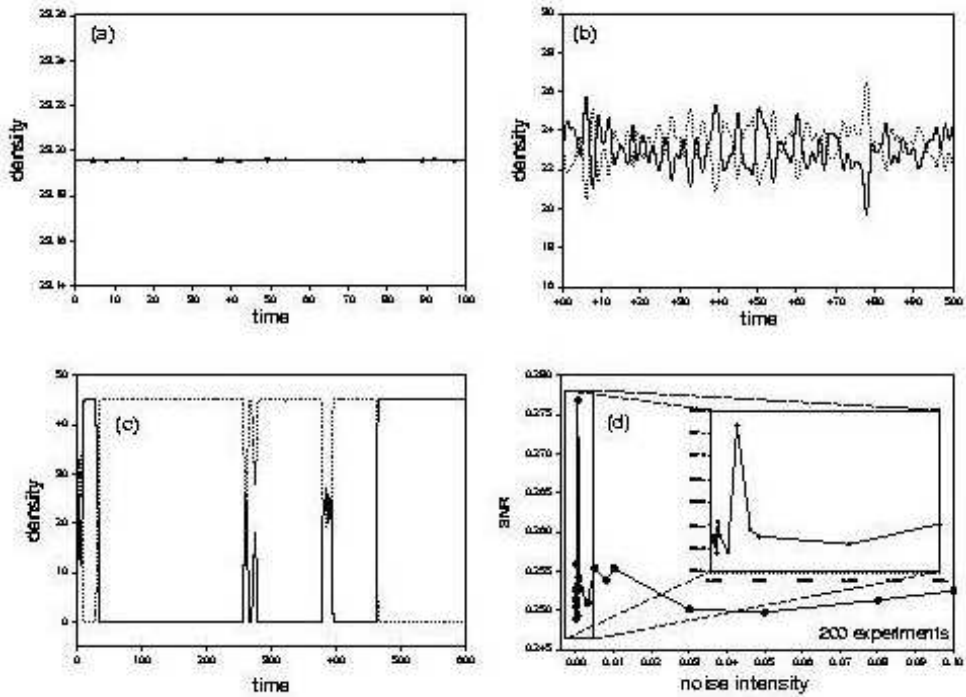}}}
%\centering{\frame{\rule{2cm}{2cm}}}
\caption{Time evolution of both population densities $x$ and $y$,
at different levels of the additive noise of Eq.(3) and SNR of
$(x-y)^2$. Species $x$ (solid line) and species $y$ (dotted line).
The values of the parameters are $\mu = 1$, $\alpha = 45$,
$\gamma=10^{-5}$, $\w0/(2\pi) = 0.34$,
$\sigma=\sigma_x=\sigma_y=10^{-9}$, and the initial values of the
two species densities are $x(0)=y(0)=1$. (a) For
$\sigma_\beta=10^{-12}$ no fluctuations of the two species
densities are observed. (b) For $\sigma_\beta$ =~$3\cdot 10^{-4}$
quasi-deterministic anticorrelated oscillations appear in the
temporal series of the two species densities. (c) For
$\sigma_\beta$ =~$10^{-3}$ a loss of coherence of the temporal
series for the two species is observed. (d) SNR corresponding to
the squared difference of population densities $(x - y)^2$.}
\end{figure}
We fix the intensity of the multiplicative noise at a constant
value $\sigma=\sigma_x=\sigma_y=10^{-9}$, and we vary the
intensity of the additive noise. For low noise intensity the two
species coexist all the time, while by increasing noise one
species dominates the other one for a random time and
quasi-deterministic oscillations take place. By increasing even
more the noise intensity the coherent response to environment
variations is lost (see Figs.2(a),(b) and (c)). These time
variations of competing populations are due to noise induced
phenomena, namely quasi-deterministic oscillations and stochastic
resonance. In Fig.2(d) we report the SNR of the squared difference
of population densities $(x-y)^2$. At a finite noise intensity
$\sigma_{\beta} = 5\cdot 10^{-4}$ the SNR exhibits a maximum,
which characterizes the SR phenomenon. Now we consider the effect
of the additive noise $\sigma_\beta $ in Eq.(\ref{beta_eq}) on the
extinction time of the species and we neglect the multiplicative
noise and the periodical forcing. Because of the initial
condition, the ecosystem is in the coexistence region, then the
deterministic extinction time of both species is infinity. By
introducing noise, exclusion takes place and a finite mean
extinction time (MET) appears. By increasing the noise intensity
we obtain a noise delayed extinction: both species survive because
of the noise. Then the average extinction time grows reaching a
saturation value, which corresponds to the absence of the barrier
of the potential $U(\beta)$. The MET as a function of the noise
intensity $\sigma_\beta $ takes a minimum at a finite noise
intensity $\sigma_{\beta}\approx 7.5 \cdot 10^{-4}$. We perform
1000 realizations of Eqs.~(\ref{LotVol}) and the results are shown
in Fig.3.

\begin{figure}[htbp]
\centering{\resizebox{9cm}{!}{\includegraphics{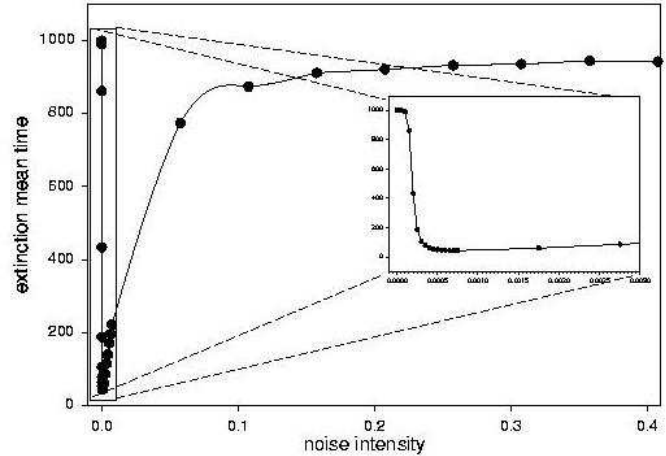}}}
%\centering{\frame{\rule{2cm}{2cm}}}
\caption{Mean extinction time of one species as a function of the
noise intensity $\sigma_\beta$ of Eq.(3). The potential $U(\beta)$
and the initial condition for $\beta(t)$ are the same of Fig.1.}
\end{figure}

\section{Spatio-Temporal Patterns}

In this section we study the effects of the noise on the spatial
distribution of interacting species. In particular we analyze the
spatial distribution of two different ecosystems: (a) two
competing species, (b) three interacting species, one predator and
two preys. To study the spatial effects we consider a discrete
time evolution model, which is the discrete version of Eqs.(1) and
(2) with diffusive terms, namely a coupled map
lattice~\cite{Spa1}.

\subsection{Two competing species}

The time evolution of the spatial distribution for the two species
is given by the following equations

\begin{eqnarray}
x_{i,j}^{n+1}&=&\mu x_{i,j}^n (1-x_{i,j}^n-\beta^n
y_{i,j}^n)+\sqrt{\sigma_x}
x_{i,j}^n X_{i,j}^n + D\sum_\gamma (x_{\gamma}^n-x_{i,j}^n),\\
y_{i,j}^{n+1}&=&\mu y_{i,j}^n (1-y_{i,j}^n-\beta^n
x_{i,j}^n)+\sqrt{\sigma_y} y_{i,j}^n Y_{i,j}^n + D\sum_\gamma
(y_{\gamma}^n-y_{i,j}^n), \label{two}
\end{eqnarray}
where $x_{i,j}^n$ and $y_{i,j}^n$ are the densities of the species
in the site \textit{(i,j)} at the time step \textit{n}, $\beta^n$
is the interaction parameter at the same time, \textit{D} is the
diffusion constant and $\sum_\gamma $ indicates the sum over the
four nearest neighbors. The $X_{i,j}^n$ and $Y_{i,j}^n$ terms are
independent Gaussian random variables with zero mean and variance
unit. The interaction parameter $\beta^n $ is a stochastic process
obtained by solving iteratively Eq.(\ref{beta_eq}). The bistable potential 
is centered at $\beta=1$ and $\beta(0) = 0.95$, which corresponds to
the minimum of the left well. The results of our simulations are shown in Fig. 4.\\
\begin{figure}[htbp]
\centering{\resizebox{12cm}{!}{\includegraphics{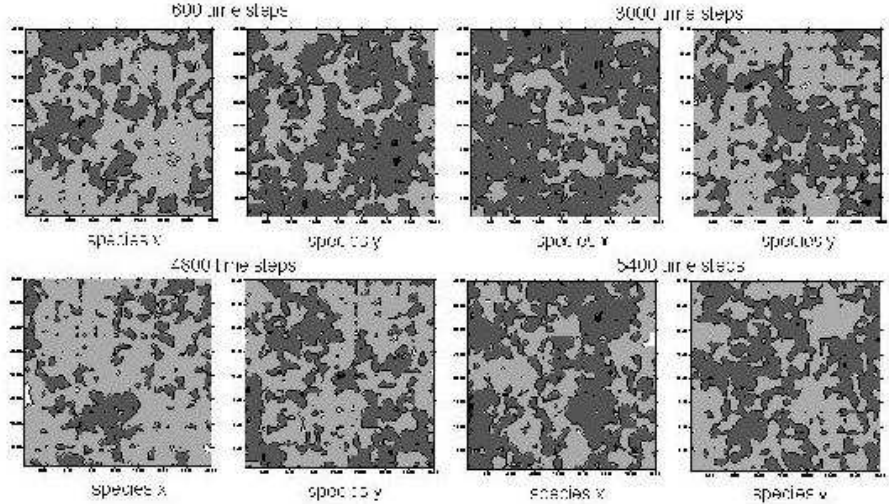}}}
%\centering{\frame{\rule{2cm}{2cm}}}
\caption{Spatio-temporal patterns induced by the noise for two
competing species. We set $\mu = 2$, $\nu = (\omega_0/2\pi) =
0.34$, $\alpha=10^{-5}$, $\sigma_x = \sigma_y = 10^{-8}$,
$\sigma_\beta = 10^{-12},~\beta(0) =0.95, D = 0.05$. The values of
the other parameters are the same used in Fig.2. The initial
spatial distribution is homogeneous and equal for the two species,
that is $x^{init}_{i,j}=y^{init}_{i,j}=0.5$ for all sites
$(i,j)$.}
\end{figure}
 We observe periodical formation of
correlated patterns. After 600 time steps the spatial
distributions of the two species appear almost independent. After
 3000 time steps the anticorrelation between the
two species densities increases. A further evolution of the system
produces again spatial distributions which are independent after
4800 time steps and anticorrelated after 5400 time steps. The two
competing species tend to occupy different spatial places
periodically without overlap.

\subsection{Three interacting species: two preys and one predator}

Finally we analyzed an ecosystem composed by three species: two
preys and one predator. We use the same coupled map lattice model
of the previous section

\begin{eqnarray}
x_{i,j}^{n+1}&=&\mu x_{i,j}^n (1-x_{i,j}^n-\beta^n
y_{i,j}^n-\gamma z_{i,j}^n)+\sqrt{\sigma_x}
x_{i,j}^n X_{i,j}^n + D\sum_\delta (x_{\delta}^n-x_{i,j}^n),\\
y_{i,j}^{n+1}&=&\mu y_{i,j}^n (1-y_{i,j}^n-\beta^n
x_{i,j}^n-\gamma z_{i,j}^n)+\sqrt{\sigma_y} y_{i,j}^n Y_{i,j}^n
+D\sum_\delta (y_{\delta}^n-y_{i,j}^n),\\
z_{i,j}^{n+1}&=&z_{i,j}^n [-\beta_z+\gamma_z(x_{i,j}^n+y_{i,j}^n)]
+ \sqrt{\sigma_z}z_{i,j}^n Z_{i,j}^n + D\sum_\delta
(z_{\delta}^n-z_{i,j}^n),\label{}
\end{eqnarray}
where $x_{i,j}^n$, $y_{i,j}^n$ and $z_{i,j}^n$ are respectively
the densities of preys \textit{x}, \textit{y} and of the predator
\textit{z} in the site \textit{(i,j)} at the time step \textit{n},
$\gamma$ and $\gamma_z$ are the interaction parameters between
preys and predator and D is the diffusion coefficient. The
interaction parameter $\beta$ between the two preys is a
periodical function whose value, after \textit{n} time steps, is
given by

\begin{equation}
\beta(t)=1+\epsilon+\alpha cos(\w0 t), \label{beta(t)}
\end{equation}
with $\epsilon=-0.01$, $\alpha=0.1$ and $\nu_0 = (\w0/2\pi) =
10^{-3}$. We consider two different initial conditions: (i) a
homogeneous initial distribution  and (ii) a peaked initial
distribution. In the first case we find exactly anticorrelated
spatial patterns of the two preys,  while the spatial patterns of
the predator show correlations with both the spatial distributions
of the preys (see Fig. 5).
\begin{figure}[htbp]
\centering{\resizebox{12cm}{!}{\includegraphics{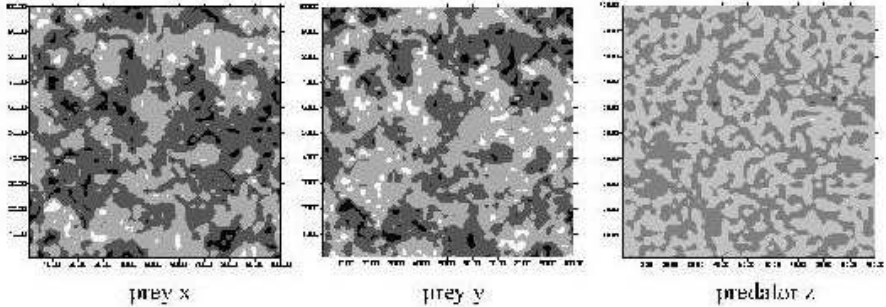}}}
%\centering{\frame{\rule{2cm}{2cm}}}
\caption{Spatial patterns induced by the noise for three
interacting species (two preys and one predator) with homogeneous
initial distributions. The parameter set is: $\epsilon=-0.01$,
$\mu = 2$, $\beta_z = 0.01$, $\nu_0 = (\w0/2\pi) = 10^{-3}$,
$\alpha=0.1$, $\sigma_x = \sigma_y = \sigma_z = 10^{-8}$, $D =
0.01, \gamma = 3\cdot10^{-2}$, $\gamma_z = 2.05 \cdot 10^2$. The
initial values of the uniform spatial distribution are
$x^{init}_{i,j} = y^{init}_{i,j}= 0.25$ and $z^{init}_{i,j} =
0.10$ for all sites \textit{(i,j)}.}
\end{figure}
The preys tend to occupy different positions as in the case of two
competing species. In the second case we use delta-like initial
distributions for the two preys and a homogeneous distribution for
the predator. After $800$ steps we find strongly correlated
spatial patterns of the preys which almost overlap each other. The
maximum of spatial distribution of the predator is just at the
boundary of the spatial concentrations of the preys, so that the
predator surrounds the preys (see Fig.6). The preys now tend to
overlap spatially as it occurs in real ecosystems when preys tend
to defend themselves against the predator attacks.

\begin{figure}[htbp]
\centering{\resizebox{12cm}{!}{\includegraphics{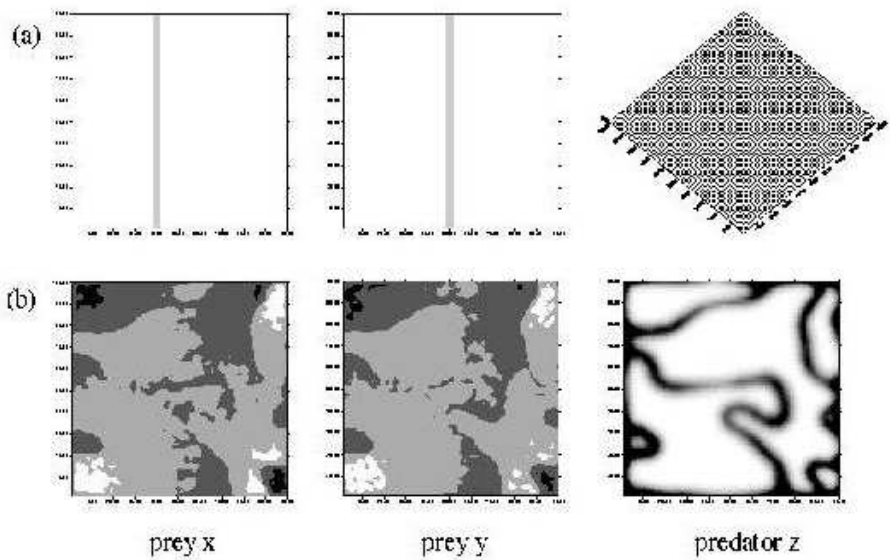}}}
%\centering{\frame{\rule{2cm}{2cm}}}
\caption{Spatial patterns induced by the noise for three
interacting species (two preys and one predator) with delta-like
initial distributions of the preys and a homogeneous distribution
of the predator: (a) initial conditions, (b) spatial patterns
after 800 time steps. Here we set $\epsilon=-0.05$, $D = 0.1$,
$\sigma_x = \sigma_y = \sigma_z = 10^{-3}$ and the other
parameters are the same as in Fig.5.}
\end{figure}
\section{Conclusions}
We analyzed the role of the noise on the spatio-temporal behaviors
of two ecosystems composed by two and three interacting species.
We use Lotka-Volterra generalized equations with multiplicative
noise, climatic periodic forcing and random interaction parameter.
The main conclusions of this work are: (i) in the case of two
interacting species the appearance of noise induced phenomena such
as quasi-deterministic oscillations, stochastic resonance, noise
delayed extinction and spatial patterns; (ii) in the case of three
interacting species the formation of dynamical spatial patterns
exhibiting correlations which are strongly dependent on the
initial conditions. Our population dynamical model could be useful
to explain spatio-temporal experimental data of interacting
species strongly interacting with the noisy environment
\cite{Gar,Car,Spr}. A more detailed study of the dynamics of
spatial patterns through stochastic partial differential equations
will be the subject of future investigations.

\section{Acknowledgements}
This work has been supported by INTAS Grant 01-450, by INFM and
MIUR.

\end{document}